\documentclass[conference]{IEEEtran}
\usepackage{cite}

\ifCLASSINFOpdf
\usepackage[pdftex]{graphicx}
\else
\fi
\usepackage{amsmath}
\usepackage{amsfonts} 
\usepackage{amsthm}
\usepackage{amssymb}

\usepackage{BOONDOX-ds}

\usepackage{bbm}

\usepackage{kbordermatrix}

\newcommand{\arrayprod}[2]{\mathbin{{ #1 }.{ #2 }}}

\newcommand{\pad}{\operatorname{pad}}

\newcommand{\sub}[3]{{\operatorname{Sub}_{#3}\left({#1},{#2}\right)}}
\newcommand{\Sub}[3]{{\operatorname{Sub}_{#3}\left({#1},{#2}\right)}}

\newcommand{\pfun}{\mathop{\hbox{$\to$\kern-7pt\raise.9pt\hbox{\scalebox{1}[.55]{$|$}}\kern4pt} }}

\newtheorem{thm}{Theorem}[section]
\newtheorem{lem}[thm]{Lemma}
\newtheorem{prop}[thm]{Proposition}

\newtheorem{definition}[thm]{Definition}

\begin{document}

\title{Polystore Mathematics of Relational Algebra}

\author{\IEEEauthorblockN{
  Hayden Jananthan$^{1,2}$,
  Ziqi Zhou$^2$,
  Vijay Gadepally$^2$,
  Dylan Hutchison$^4$,
  Suna Kim$^{2,3}$,
  Jeremy Kepner$^{2}$
}
\vspace{1ex}
\IEEEauthorblockA{
	$^1$Vanderbilt University,
	$^2$MIT,
	$^3$CalTech,
	$^4$University of Washington
}
}
\maketitle

\begin{abstract}
Financial transactions, internet search, and data analysis are all placing increasing demands on databases.  SQL, NoSQL, and NewSQL databases have been developed to meet these demands and each offers unique benefits.  SQL, NoSQL, and NewSQL databases also rely on different underlying mathematical models.  Polystores seek to provide a mechanism to allow applications to transparently achieve the benefits of diverse databases while insulating applications from the details of these databases.  Integrating the underlying mathematics of these diverse databases can be an important enabler for polystores as it enables effective reasoning across different databases.  Associative arrays provide a common approach for the  mathematics of polystores by encompassing the mathematics found in different databases: sets (SQL), graphs (NoSQL), and matrices (NewSQL).  Prior work presented the SQL relational model in terms of associative arrays and identified key mathematical properties that are  preserved within SQL. This work provides the rigorous mathematical definitions, lemmas, and theorems underlying these properties.  Specifically, SQL Relational Algebra deals primarily with relations -- multisets of tuples -- and operations on and between those relations. These relations can be modeled as associative arrays by treating tuples as non-zero rows in an array. Operations in relational algebra are built as compositions of standard operations on associative arrays which mirror their matrix counterparts. These constructions provide insight into how relational algebra can be handled via array operations. As an example application, the composition of two projection operations is shown to also be a projection, and the projection of a union is shown to be equal to the union of the projections.
\end{abstract}

%
\IEEEpeerreviewmaketitle

\section{Introduction}
\let\thefootnote\relax\footnotetext{This material is based in part upon
work supported by the NSF under grant number DMS-1312831.  Any opinions,
findings, and conclusions or recommendations expressed in this material
are those of the authors and do not necessarily reflect the views of
the National Science Foundation.}
The success of SQL, NoSQL, and NewSQL databases is a reflection of their ability to provide significant functionality and performance benefits for specific domains, such as financial transactions, internet search, data analysis, and, increasingly, machine learning.  Polystore databases seek to provide a mechanism to allow applications to transparently achieve the benefits of diverse databases while insulating applications from the details of these databases.  Polystores must support a wide range of databases with different iterfaces.  Among these interfaces are the standard Relational or SQL (Structured Query Language) databases \cite{Codd1970, Stonebraker1976} such as MySQL, PostgreSQL, and Oracle; key-value stores/NoSQL databases such as Google BigTable \cite{Chang2008}, Apache Accumulo \cite{Wall2015}, and MongoDB \cite{Chodorow2013}; NewSQL databases such as C-Store \cite{Stonebraker2005}, H-Store \cite{Kallman2008}, SciDB \cite{Balazinska2009}, VoltDB \cite{StonebrakerWeisberg2013}, and Graphulo \cite{Hutchison2015,gadepally2015graphulo}. 

NoSQL databases were developed to represent large sparse tables, contributing to the widespread adoption of NoSQL databases to analyze data on the internet \cite{DeCandia2007, LakshmanMalik2010, George2011}. NewSQL databases support new analytics capabilities within a database.  In hybrid processing systems like Apache Pig \cite{Olston2008}, Apache Spark \cite{Zaharia2010}, and HaLoop \cite{Bu2010}, SQL, NoSQL, and NewSQL concepts have been blended.

\begin{figure}[h] 
\centering
\includegraphics[scale=1]{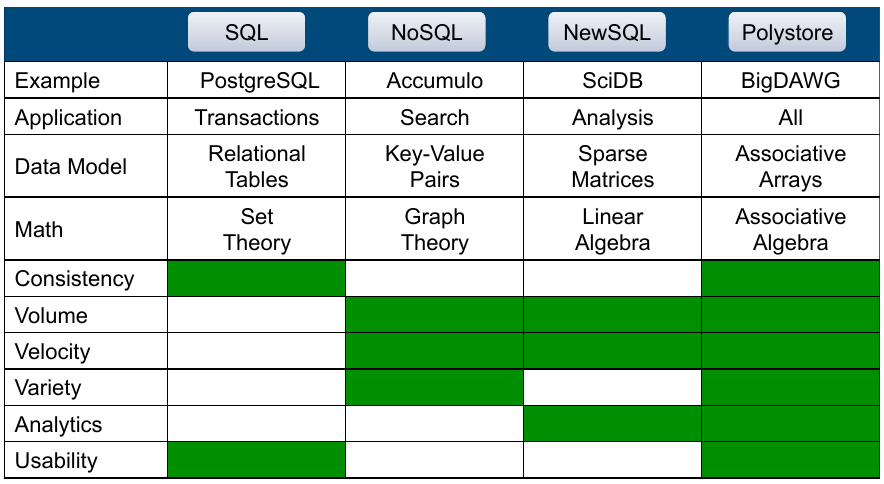}
\caption{Focus areas of SQL, NoSQL, NewSQL, and Polystore databases.  Each class of database has distinct strengths and relies on a different mathematics.  Polystores provide a way to unify these databases and their mathematics.}
\label{focus areas figure}
\end{figure}

\begin{figure}[h] 
\centering
\includegraphics[scale=1]{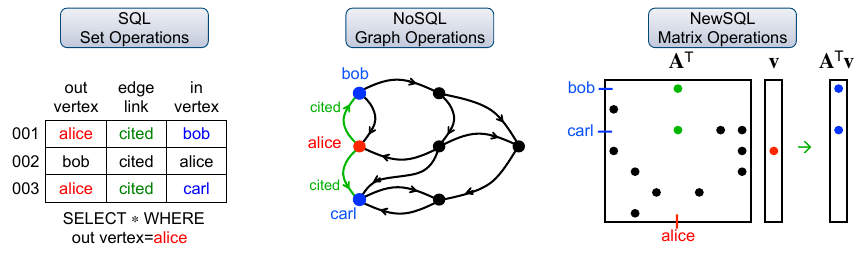}
\caption{Breadth-first search for SQL, NoSQL, and NewSQL databases in their native data models.  In each case, the operation being performed is finding the nearest neighbors of {\sf alice} who are {\sf bob} and {\sf carl}.}
\label{bfs figure}
\end{figure}

Polystore databases, such as BigDAWG \cite{Duggan2015, Elmore2015, gadepally2016bigdawg, gadepally2017version, obrien2017bigdawg} and Myria \cite{wang2016myriaOverview}, were created to make use of the varied specialties of the aforementioned database types \cite{StonebrakerCetintemel2005}. One inherent challenge is that SQL, NoSQL, and NewSQL databases use different data models and make use of different mathematical tools, as illustrated by Figure~\ref{focus areas figure} and Figure~\ref{bfs figure}.

Integrating the mathematics of these diverse databases is an important enabler for polystores as it allows reasoning across different databases.  
The mathematical foundations for these databases include sets (SQL), graphs (NoSQL), and matrices (NewSQL).
\textsc{Lara} \cite{hutchison2017laradb} is one branch of work that reduces the mathematics of sets (Relational Algebra) and matrices (Linear Algebra) to a common basis of 3 operators, emphasizing their practical realization in data processing systems.
This paper focuses on associative arrays as a common approach for the mathematics of polystores by situating Relational Algebra onto the same foundations as Linear Algebra.

The associative array approach to bridging NoSQL and NewSQL databases has been demonstrated by the Dynamic Distributed Dimensional Data Model (D4M) technology \cite{Kepner2012} which provides a linear algebraic interface to SQL, NoSQL, and NewSQL databases \cite{Byun2012, Kepner2013, gadepally2015d4m, chen2016julia, milechin2017d4m}. The key object of D4M is the \emph{associative array}, which generalizes the notion of a matrix to allow for more general value sets and indexing sets. This more general structure makes associative arrays better equipped to deal with graphs and relations in a more direct fashion than matrices \cite{KepnerChaidez2013, KepnerChaidez2014, jananthan2017constructing, kepnerjananthan}.

Relations form the basis of the mathematical foundation for SQL databases \cite{Maier1983, Codd1990, Abiteboul1995}. In set theory, they are realized as multisets of tuples. Associative arrays can be used to realize multisets of tuples of values which support a notion of ``addition'' and ``multiplication''. Associative array analogues of traditional matrix operations can be defined, and this leads to the question of whether the gamut of relational algebra operations can be likewise realized in terms of the associative array operations.

Our prior work \cite{kepner2016} presented the SQL relational model in terms of associative arrays and identified key mathematical properties that are  preserved within SQL. This work provides the rigorous mathematical definitions and proofs of some of these properties.  Specifically, SQL Relational Algebra deals primarily with relations -- multisets of tuples -- and operations on and between those relations. These relations can be modeled as associative arrays by treating tuples as non-zero rows in an array. Operations in relational algebra can be built as compositions of standard operations on associative arrays which mirror their matrix counterparts. These constructions provide insight into how relational algebra can be handled via array operations.

This paper gives some technical background of multisets and associative arrays to explain the motivation for identifying relations and associative arrays, defines some major relational algebra operations in terms of standard associative array operations, and uses these definitions to prove fundamental properties of some of the relational algebra operations.

\section{Mathematical Preliminaries and Definitions}

Understanding relations in terms of associative arrays begins with the careful definition of the relevant mathematical properties of relations and associative arrays.

\subsection{Relations and Multisets in Set Theory}

Relations form the basis of the mathematical foundation for SQL databases \cite{Maier1983, Codd1990, Abiteboul1995}. In set theory, they are realized as multisets of tuples.

One approach to defining multisets in set theory that matches intuition closely is to define a \emph{multiset} as a sequence
\begin{equation*}
  f: I \to A
\end{equation*}
in which $f^{-1}(a)$ is a non-empty finite set for each $a\in A$. The size of $f^{-1}(a)$ represents how many copies of $a$ are in the multiset. Two sequences
\begin{equation*}
  f: I \to A \quad \text{and} \quad  g: J \to B
\end{equation*}
are said to \emph{define the same multiset} if $A=B$ and there exists a bijection
\[
  h: I \to J
\] such that
\[
  g\circ h = f
\]
This definition of equality of multisets captures the fact that the specific indexing set $I$ used does not matter, only the sizes of $f^{-1}(a)$ (which is invariant under equality of multisets).

This makes sense of something like $\{a,b,a,c,e,a,b\}$ as the multiset
\[
  f: \{1,\ldots,7\} \to \{a,b,c,e\}
\]
where $f(1) = a$, $f(2)=b$, $f(3)=a$, $f(4) = c$, $f(5)=e$, $f(6)=a$, $f(7)=b$,
In this notation, $\{a,a,a,b,b,c,e\}$ defines the same multiset.

A slight variation on this definition is to allow $A$ to contain non-elements, i.e. an $a\in A$ such that $f^{-1}(a)$ is empty. Accepting this variation makes the equivalance with associative arrays technically simpler.

In relational algebra, a relation is a multiset of tuples. 
These tuples conform to a schema of $n$ attributes, where $n$ is the arity of the relation.
For example, an arity-3 relation might contain the tuple (7, ``Hayden'', 20).
We assume all relations contain a primary key attribute; if not, then a primary key can be appended to the relation as an additional attribute, as many databases do in practice under the hood.
If $J$ refers to the relation's primary key and $\mathbb{V}$ refers to (the cross product of) its other attributes' domains, then we can define a relation as
\[
  f: J \to \mathbb{V}
\]
which maps the primary key value of a tuple to its other attributes' values. For example, such a mapping might contain the entry 7 $\rightarrow$ (``Hayden'', 20)
Primary keys outside the relation map to all-null rows, as discussed in the next section.

Note that while both multisets and tuples are of the form
\[
  f: I \to A
\]
they differ in that the specific set $I$ is inconsequential in the definition of a multiset, while it is important in the case of tuples.

\subsection{Associative Arrays}

In practice, the values in a tuple can range from alphanumeric strings to real numbers to sets. Moreover, the kinds of operations defined on those values need not be the traditional addition and multiplication of real numbers. However, in order to define analogues of the standard matrix operations, there need be some ``addition'' $\oplus$ and some ``multiplication'' $\otimes$, and these should satisfy some minimum set of properties to ensure that these array-analogues have a minimum set of desireable properties. 

Generalizing the notion of a matrix to allow for more general value sets equipped with more general operations produces the notion of an associative array. 

\begin{definition}[Semiring] \cite{gondran2007, golan2013}
A \emph{semiring} is a set $\mathbb{V}$ equipped with two binary operations $\oplus$ and $\otimes$ such that
\begin{enumerate}
\item $\oplus$ is associative and commutative and has an identity element $0 \in \mathbb{V}$,
\item $\otimes$ is associative with an identity element $1\in\mathbb{V}$,
\item $\otimes$ distributes over $\oplus$, and
\item $0$ is an annihilator for $\otimes$.
\end{enumerate}
\end{definition}

All rings and fields are semirings. 
The set of natural numbers $\mathbb{N} = \{0,1,2,\ldots\}$ is a semiring under standard addition and multiplication. 
The set of non-negative real numbers is a semiring under standard addition and multiplication. 
The set of extended real numbers $\mathbb{R}\cup \{-\infty,\infty\}$ with semiring addition $\oplus = \max$ and semiring multiplication $\otimes = \min$ is a semiring called the \emph{max-min algebra}. 
$\mathbb{R}\cup\{-\infty,\infty\}$ with $\oplus=\max$ and $\otimes = +$ is a semiring called the \emph{max-plus algebra}. 
The set of alphanumeric strings ordered lexicographically along with a formal maximum $\infty$ is a semiring with $\oplus = \min$ and $\otimes = \text{concatenation}$.

The convention that $\mathrm{null} = 0$ is used here. Note that if a formal $\mathrm{null}$ is added to $\mathbb{V}$ with the properties
\begin{align*}
\mathrm{null}\oplus v = v \oplus \mathrm{null} & = v \\
\mathrm{null}\otimes v = v\otimes \mathrm{null} & = \mathrm{null}
\end{align*}
for every $v\in \mathbb{V} \cup \{\mathrm{null}\}$, then $\mathbb{V} \cup \{\mathrm{null}\}$ would be a semiring with a new additive identity $\mathrm{null}$. Thus, nothing is lost by examining only the case where the convention $\mathrm{null}=0$ is used.

\begin{definition}[Associative Array]
An \emph{associative array} is a map
\[
  \mathbf{A}: K_1\times K_2 \to \mathbb{V}
\]
where $\mathbb{V}$ is a semiring, such that $\mathbf{A}(k_1,k_2) \neq 0$ for only finitely-many pairs $(k_1,k_2)$. 
Elements of $K_1$ are called \emph{row keys} and elements of $K_2$ are called \emph{column keys}.
\end{definition}

\begin{definition}[Array Addition]
Suppose
\[
  \mathbf{A},\mathbf{B}: K_1 \times K_2 \to \mathbb{V}
\]
are two associative arrays. Their \emph{array addition}
\[
  \mathbf{C} = \mathbf{A}\oplus \mathbf{B}: K_1 \times K_2 \to \mathbb{V}
\]
is defined by
\begin{equation*}
\mathbf{C}(k_1,k_2) = (\mathbf{A}\oplus \mathbf{B})(k_1,k_2) = \mathbf{A}(k_1,k_2) \oplus \mathbf{B}(k_1,k_2)
\end{equation*}
\end{definition}

\noindent Array addition is both associative and commutative.

\begin{definition}[Zero Array]
$\mathbs{0}$ is the \emph{zero array}, in which every entry is $0$.
\end{definition}

\noindent The zero array provides an identity element for array addition.

\begin{definition}[Element-Wise Product]
Suppose
\[
  \mathbf{A},\mathbf{B}: K_1 \times K_2 \to \mathbb{V}
\]
are two associative arrays. Their \emph{array element-wise product}
\[
  \mathbf{C} = \mathbf{A}\otimes \mathbf{B} : K_1 \times K_2 \to \mathbb{V}
\]
is defined by
\begin{equation*}
\mathbf{C}(k_1,k_2) = (\mathbf{A}\otimes \mathbf{B})(k_1,k_2) = \mathbf{A}(k_1,k_2) \otimes \mathbf{B}(k_1,k_2)
\end{equation*}
\end{definition}

Array element-wise product is associative, as well as commutative if $\otimes$ is commutative.

\begin{definition}[Element-Wise Identity]
Given a row key set $K_1$ and a column key set $K_2$, denote by $\mathbbm{1}_{K_1,K_2}$ the associative array $K_1\times K_2 \to \mathbb{V}$ with
\begin{equation*}
\mathbbm{1}_{K_1,K_2}(k_1,k_2) = 1
\end{equation*}
\end{definition}

\noindent The element-wise identity $\mathbbm{1}_{K_1,K_2}$ provides an identity element for array element-wise product when restricted to associative arrays $\mathbf{A}: K_1 \times K_2 \to \mathbb{V}$.

\begin{definition}[Array Multiplication] \label{array multiply definition}
Suppose
\[
  \mathbf{A}: K_1 \times K_2 \to \mathbb{V}
\]
and
\[
  \mathbf{B}: K_2 \times K_3 \to \mathbb{V}
\]
are two associative arrays. Then their \emph{array product}
\[
  \mathbf{C} = \mathbf{A}~\mathbf{B} = \mathbf{A} \arrayprod{\oplus}{\otimes} \mathbf{B}: K_1\times K_3 \to \mathbb{V}
\]
is defined by
\begin{equation*}
\mathbf{C}(k_1,k_3) = \bigoplus_{k_2 \in K_2}{\mathbf{A}(k_1,k_2)\otimes \mathbf{B}(k_2,k_3)}
\end{equation*}
\end{definition}

\noindent Array multiplication is associative, but in general need not be commutative even if $\otimes$ is commutative.

For brevity, $\mathbf{A}\arrayprod{\oplus}{\otimes} \mathbf{B}$ is denoted $\mathbf{A}~\mathbf{B}$, except when it is important to be explicit about the operations being used (particularly when they are not semiring $\oplus$ and $\otimes$).

\begin{definition}[Array Identity]
Given a row key set $K_1$, a column key set $K_2$, and a partial function
\[
f: K_1 \pfun K_2
\]
meaning $f$ is a function defined on a subset $\operatorname{dom} f \subset K_1$. Denote by
\[
  \mathbb{I}_{K_1,K_2,f}
\] the associative array $K_1\times K_2 \to \mathbb{V}$ with
\begin{equation*}
\mathbb{I}_{K_1,K_2,f}(k_1,k_2) = \begin{cases} 1 & \text{if $k_1 \in \operatorname{dom} f$ and $k_2 = f(k_1)$} \\ 0 & \text{otherwise} \end{cases}
\end{equation*}
$\mathbb{I}_{K_1,K_2}$ means $\mathbb{I}_{K_1,K_2,f}$ where
\[
  \operatorname{dom} f = K_1\cap K_2
\]
and $f$ acts as the identity on $K_1 \cap K_2$. If $K_1=K_2=K$, then
\[
  \mathbb{I}_{K_1,K_2} = \mathbb{I}_K
\]
Finally, if $K_1,K_2,f$ are understood, then write
\[
  \mathbb{I}_{K_1,K_2,f} = \mathbb{I}
\]
\end{definition}

\noindent The array identity $\mathbb{I}_{K_1,K_2,f}$ does not in general act as an identity element for array multiplication. $\mathbb{I}_K$, however, is an identity for array multiplication when restricted to associative arrays $\mathbf{A}: K\times K \to \mathbb{V}$.

When performing operations between associative arrays whose row and column key sets are not ``compatible'' (i.e. satisfy the hypotheses of the above definitions), this can be solved by zero padding.

\begin{definition}[Zero Padding]
If
\[
  \mathbf{A}: K_1 \times K_2 \to \mathbb{V}
\]
is an associative array and $K_1',K_2'$ are arbitrary sets, then 
\begin{equation*}
\pad_{K_1'\times K_2'}(\mathbf{A}): (K_1\cup K_1')\times (K_2\cup K_2') \to \mathbb{V}
\end{equation*}
is defined by
\begin{equation*}
\pad_{K_1'\times K_2'}(\mathbf{A})(i,j) = \begin{cases} \mathbf{A}(i,j) & \text{if $i\in K_1,j\in K_2$} \\ 0 & \text{otherwise} \end{cases}
\end{equation*}
\end{definition}

\noindent By padding arrays prior to carrying out an operation, operations can be defined in general. Explicitly, given
\[
  \mathbf{A}: K_1 \times K_2 \to \mathbb{V}
\]
and
\[
  \mathbf{B}: K_3 \times K_4 \to \mathbb{V}
\]
then
\begin{align*}
\mathbf{A} \oplus \mathbf{B} & = \pad_{(K_1 \cup K_3) \times (K_2\cup K_4)}(\mathbf{A}) \\
& \quad \quad \oplus \pad_{(K_1\cup K_3) \times (K_2 \cup K_4)}(\mathbf{B}) \\
\mathbf{A} \otimes \mathbf{B} & = \pad_{(K_1 \cup K_3) \times (K_2\cup K_4)}(\mathbf{A}) \\
& \quad \quad \otimes \pad_{(K_1\cup K_3) \times (K_2 \cup K_4)}(\mathbf{B}) \\
\mathbf{A} \arrayprod{\oplus}{\otimes} \mathbf{B} & = \pad_{K_1 \times (K_2 \cup K_3)}(\mathbf{A}) \\
& \quad \quad \arrayprod{\oplus}{\otimes} \pad_{(K_2 \cup K_3) \times K_4}(\mathbf{B})
\end{align*}
Likewise, equality of two arrays is done up to zero padding: $\mathbf{A} = \mathbf{B}$ if and only if
\begin{equation*}
\pad_{(K_1 \cup K_3) \times (K_2 \cup K_4)}(\mathbf{A}) = \pad_{(K_1\cup K_3)\times (K_2 \cup K_4)}(\mathbf{B})
\end{equation*}

\begin{definition}[Row Support]
For an associative array
\[
  \mathbf{A}: K_1 \times K_2 \to \mathbb{V}
\]
the \emph{row support} $I_\mathbf{A}$ is the set of row keys associated with non-zero rows of $\mathbf{A}$.
\end{definition}

\begin{definition}[Transpose]
Suppose
\[
  \mathbf{A}: K_1 \times K_2 \to \mathbb{V}
\]
is an associative array. Then its \emph{transpose} $\mathbf{A}^\intercal$ is the associative array $K_2 \times K_1 \to \mathbb{V}$ defined by
\begin{equation*}
\mathbf{A}^\intercal(k_2,k_1) = \mathbf{A}(k_1,k_2)
\end{equation*}
\end{definition}

\begin{definition}[Array Kronecker Product]
Suppose
\[
  \mathbf{A}: K_1 \times K_2 \to \mathbb{V}
\]
and
\[
  \mathbf{B}: K_3\times K_4 \to \mathbb{V}
\]
are associative arrays. Then their \emph{array Kronecker product}
\[
 \mathbf{C} = \mathbf{A}
   \mathbin{\text{\textcircled{$\otimes$}}}
  \mathbf{C}
\]
is the associative array
\[
  (K_1 \times K_3) \times (K_2 \times K_4) \to \mathbb{V}
\]
defined by
\begin{equation*}
\mathbf{C}((k_1,k_3),(k_2,k_4)) = \mathbf{A}(k_1,k_2) \otimes \mathbf{B}(k_3,k_4)
\end{equation*}
\end{definition}

\noindent The array Kronecker product allows associative arrays operations to handle dimensions higher than $2$ dimensions.

\section{Relations as Associative Arrays}

Motivated by the definition of a relation as a multiset (sequence) of tuples, \emph{define} a relation to be an associative array with the intuition that the rows of an associative array are the relevant tuples which are indexed by the column indices. The row indices are only meant to differentiate the rows. In practice, the sequence number identifying the distinct rows in an SQL table serve a similar purpose.

\begin{definition}[Row and Row Equality]
If
\[
  \mathbf{A}: K_1 \times K_2 \to \mathbb{V}
\]
is an associative array with $i\in K_1$, then the \emph{$i$-th row} is the tuple
\[
  \mathbf{A}(i,:) : K_2 \to \mathbb{V}
\]
sending $j$ to $\mathbf{A}(i,j)$. Such a row is \emph{non-zero} if it not identically zero.
If
\[
  \mathbf{B}: K_3 \times K_4 \to \mathbb{V}
\]
and $i' \in K_3$, then the $i$-th row of $\mathbf{A}$ is \emph{equal to} the $i'$-th row of $\mathbf{B}$ if $\mathbf{A}(i,j)$ and $\mathbf{B}(i',j)$ are both defined and equal whenever one of them is non-zero.
$\mathbf{A}^{-1}(i,:)$ denotes the subset of $I_\mathbf{A}$ containing the indices of rows in $\mathbf{A}$ which are equal to $\mathbf{A}(i,:)$.
\end{definition}

\begin{definition}[Weak Equivalence]
The associative arrays
\begin{equation*}
  \mathbf{A}: K_1\times K_2 \to \mathbb{V}
\end{equation*}
and
\begin{equation*}
  \mathbf{B}:K_3\times K_4 \to \mathbb{V}
\end{equation*}
are \emph{weakly equivalent}
\begin{equation*}
  \mathbf{A} \sim \mathbf{B}
\end{equation*}
if for each non-zero row of $\mathbf{A}$ there is an equal row in $\mathbf{B}$, and vice-a-versa.
\end{definition}

\noindent In terms of multisets, two arrays are weakly equivalent if their underlying sets of tuples are equal.

\begin{lem} \label{weak equivalence in terms of cross-correlation matrix}
Given associative arrays
\[
  \mathbf{A}: K_1 \times K_2 \to \mathbb{V}
\]
and
\[
  \mathbf{B}: K_3 \times K_4 \to \mathbb{V}
\]
Define the array
\[
  \mathbf{P}: K_1 \times K_3 \to \mathbb{V}
\]
by
\begin{equation*}
\mathbf{P}(k_1,k_3) = \begin{cases} 1 & \text{ if $\mathbf{A}(k_1,:) = \mathbf{B}(k_3,:)$} \\ 0 & \text{otherwise} \end{cases}
\end{equation*}
Then the following are equivalent
\begin{enumerate}
\item $\mathbf{A}\sim \mathbf{B}$.
\item If $\mathbf{A}(k_1,:)$ is a non-zero row, so is the row $\mathbf{P}(k_1,:)$, and if $\mathbf{B}(k_3,:)$ is a non-zero row, so is the column $\mathbf{P}(:,k_3)$.
\end{enumerate}
\end{lem}

\begin{lem}
$\mathbf{A} \sim \mathbf{B}$ if and only if there exist functions
\[
  f: I_\mathbf{A} \to I_\mathbf{B}
\]
and
\[
  g: I_\mathbf{B} \to I_\mathbf{A}
\]
such that
\begin{equation*}
\mathbf{A} = \mathbb{I}_{I_\mathbf{A},I_\mathbf{B},f} ~
\mathbf{B} \quad \text{and} \quad \mathbf{B} = \mathbb{I}_{I_\mathbf{B},I_\mathbf{A},g} ~
\mathbf{A} 
\end{equation*}
\end{lem}

\begin{definition}[Strong Equivalence]
Two associative arrays
\[
  \mathbf{A}: K_1 \times K_2 \to \mathbb{V}
\]
and
\[
  \mathbf{B}: K_3 \times K_4 \to \mathbb{V}
\]
are strongly equivalent
\[
  \mathbf{A} \approx \mathbf{B}
\]
if for each non-zero row of $\mathbf{A}$, there are exactly as many copies of that row in $\mathbf{B}$ as in $\mathbf{A}$, and vice-a-versa.
\end{definition}

\noindent In terms of multisets, two arrays are strongly equivalent if they are equal as multisets.

\begin{lem} \label{strong equivalence in terms of cross-correlation matrix}
For associative arrays
\[
  \mathbf{A}: K_1 \times K_2 \to \mathbb{V}
\]
and
\[
  \mathbf{B}: K_3 \times K_4 \to \mathbb{V}
\]
define the array
\[
  \mathbf{P}: K_1 \times K_3 \to \mathbb{V}
\]
by
\begin{equation*}
\mathbf{P}(k_1,k_3) = \begin{cases} 1 & \text{ if $\mathbf{A}(k_1,:) = \mathbf{B}(k_3,:)$} \\ 0 & \text{otherwise} \end{cases}
\end{equation*}
Then the following are equivalent:
\begin{enumerate}
\item $\mathbf{A} \approx \mathbf{B}$
\item $\mathbf{A} \sim \mathbf{B}$ and if $\mathbf{P}(k_1,k_3) \neq 0$, then the number of non-zero entries of the row $\mathbf{P}(k_1,:)$ is equal to the number of non-zero entries of the column $\mathbf{P}(:,K_3)$.
\end{enumerate}
\end{lem}

\begin{lem}
$\mathbf{A}\approx \mathbf{B}$ if and only if there exists a bijection
\[
  f: I_\mathbf{A} \to I_\mathbf{B}
\]
such that
\begin{equation*}
\mathbf{A} = \mathbb{I}_{I_\mathbf{A},I_\mathbf{B},f} ~
\mathbf{B}
\end{equation*}
\end{lem}

\noindent The array $\mathbf{P}$ constructed in \ref{weak equivalence in terms of cross-correlation matrix} and \ref{strong equivalence in terms of cross-correlation matrix} can be computed using the following array operation.

\begin{lem}
For associative arrays
\[
  \mathbf{A}: K_1 \times K_2 \to \mathbb{V}
\]
and
\[
  \mathbf{B}: K_3 \times K_4 \to \mathbb{V}
\]
define the array
\[
  \mathbf{P}: K_1 \times K_3 \to \mathbb{V}
\]
by
\begin{equation*}
\mathbf{P}(k_1,k_3) = \begin{cases} 1 & \text{ if $\mathbf{A}(k_1,:) = \mathbf{B}(k_3,:)$} \\ 0 & \text{otherwise} \end{cases}
\end{equation*}
Then
\begin{equation*}
\mathbf{P} = (\mathbb{I}_{I_\mathbf{A}} ~
\mathbf{A}) \arrayprod{\wedge}{\delta}
(\mathbb{I}_{I_\mathbf{B}} ~
\mathbf{B})^\intercal
\end{equation*}
where
\begin{equation*}
v \wedge w = \begin{cases} 1 & \text{ if $v,w \neq 0$} \\ 0 & \text{otherwise} \end{cases} \quad \delta(v,w) = \begin{cases} 1 & \text{if $v=w$} \\ 0 & \text{otherwise} \end{cases}
\end{equation*}
\end{lem}

\section{Relational Algebra Operations}

There are many operations defined on relations.  This is a result of of Codd's Theorem \cite{Codd1972}, which states that relational algebra and relational calculus queries have the same expressive power. In other words, to carry out a wide range of relational queries, it is enough to implement the basic relational algebra operations. By implementing these operations with associative algebra operations, this reduces SQL queries to linear algebra.

Equipped with the necessary array operations and notions of equivalence for arrays viewed as relations, it is possible to define several of the standard operations in relational algebra in terms of array operations.

\begin{definition}[Project] \label{project definition}
Suppose $J$ is a set of column keys.  Then define the \emph{projection operation} $\Pi_J$ by
\begin{equation*}
\Pi_J(\mathbf{A}) = \mathbf{A} ~
\mathbb{I}_J
\end{equation*}
which removes all columns not in $J$.
In SQL syntax, it is 
\begin{equation*}
\text{\sf SELECT $J(1),\ldots, J(n)$ FROM $\mathbf{A}$}
\end{equation*}
\end{definition}

\begin{definition}[Rename] \label{rename definition}
Suppose $J_1,J_2$ are two sets of column keys with a bijection $f: J_1 \to J_2$. Then define the \emph{rename operation} $\rho_{J_1/J_2,f}$ by
\begin{equation*}
\rho_{J_1/J_2,f}(\mathbf{A}) = \mathbf{A} ~
\mathbb{I}_{J_1,J_2,f}
\end{equation*}
where
\[
  \mathbf{A}: K_1\times K_2 \to \mathbb{V}
\]
This operation selects the columns to be renamed, and then renames them according to $f$.
In SQL syntax, it is
\begin{equation*}
\text{\sf SELECT $J_1(1),\ldots, J_1(n)$ AS $J_2(1),\ldots, J_2(n)$ FROM $\mathbf{A}$}
\end{equation*}
\end{definition}

\noindent The function $f$ can act trivially on some columns. This allows the rename operation to keep fixed the columns that aren't being renamed to something new.

\begin{definition}[Union] \label{union definition}
For two arrays
\[
  \mathbf{A}: K_1\times K_2 \to \mathbb{V}
\]
and
\[
  \mathbf{B}:K_3\times K_4 \to \mathbb{V}
\]
their \emph{union} is
\begin{equation*}
\mathbf{A} \cup \mathbf{B} = (\mathbb{I}_{I_\mathbf{A} \times \{1\},I_\mathbf{A}} ~
\mathbf{A}) \oplus (\mathbb{I}_{I_\mathbf{B}\times \{2\},I_\mathbf{B}} ~ 
\mathbf{B})
\end{equation*}
This operation effectively adds the counts of rows together.
In SQL syntax, it is
\begin{equation*}
\text{\sf SELECT $\ast$ FROM $\mathbf{A}$ UNION ALL SELECT $\ast$ FROM $\mathbf{B}$}
\end{equation*}
\end{definition}

\subsection{Operations Involving Choices of Representative Rows}

\noindent In the definition of a multiset intersection operation, the number of times a row appears in $\mathbf{A} \cap \mathbf{B}$ should be the minimum of the number of times that row appears in both $\mathbf{A}$ and $\mathbf{B}$. Similarly, in the definition of a multiset difference operation, the number of times a row appears in $\mathbf{A}\setminus \mathbf{B}$ should be the number of times it appears in $\mathbf{A}$ minus the number of times it appears in $\mathbf{B}$ (showing up zero times if this difference is negative). This suggests a difficult arising due to needed to select the relevant number of rows, which arises due to explictly having these rows indexed in a way that may not offer an unambiguous way of making this choice. Assume there is a function
\[
  \sub{A}{B}{n}
\]
which assigns to any sets of row keys $A$ and $B$ and a non-negative integer
\[
  0 \leq n \leq |A|+|B|
\]
a fixed subset of
\[
  (A\times \{1\}) \cup (B\times \{2\})
\]
of size $n$.

If there is a fixed, explicit total ordering of the row keys, then
\[
  (A\times \{1\}) \cup (B\times \{2\})
\]
has a canonical ordering coming from the ordering of the rows; if
\[
  A = \{a_1,\ldots, a_n\}
\]
with $a_1 < \cdots < a_n$ and
\[
  B = \{b_1,\ldots, b_m\}
\]
with $b_1 < \cdots < b_m$ then
\begin{equation*}
(a_1,1) < \cdots < (a_n,1) < (b_1,2) < \cdots < (b_m,2)
\end{equation*}
Then $\sub{A}{B}{n}$ can be taken to be the first $n$ elements of
\[
  (A\times \{1\}) \cup (B\times \{2\})
\]
with respect to the above ordering.

\begin{definition}[Intersection] \label{intersection definition}
The \emph{intersection operation} is defined by
\begin{equation*}
\mathbf{A} \cap \mathbf{B} = \mathbb{I}_S ~
(\mathbf{A} \cup \mathbf{B})
\end{equation*}
where 
\begin{equation*}
S = \bigcup_{\substack{i_1\in I_\mathbf{A}\\ i_2 \in I_\mathbf{B} \\ \mathbf{A}(i_1,:) = \mathbf{B}(i_2,:)}}{\sub{\mathbf{A}^{-1}(i_1,:)}{\mathbf{B}^{-1}(i_2,:)}{\min(m_{i_1},n_{i_2})}}
\end{equation*}
where
\[
  m_{i_1} = |\mathbf{A}^{-1}(i_1,:)|
\]
and
\[
  n_{i_2} = |\mathbf{B}^{-1}(i_2,:)|
\]
This selects from $\mathbf{A} \cup \mathbf{B}$ the minimum of the count of each row from $\mathbf{A}$ and $\mathbf{B}$.
In SQL syntax, it is
\begin{equation*}
\text{\sf SELECT $\ast$ FROM $\mathbf{A}$ INTERSECT SELECT $\ast$ FROM $\mathbf{B}$}
\end{equation*}
\end{definition}

\begin{definition}[Multiset Difference] \label{set difference definition}
The \emph{multiset difference operation} is defined by
\begin{equation*}
\mathbf{A}\setminus \mathbf{B} = \mathbb{I}_S ~
\mathbf{A}
\end{equation*}
where
\begin{equation*}
S = I_\mathbf{A} \setminus \pi_1\left[ \bigcup_{\substack{i_1\in I_\mathbf{A} \\ i_2 \in I_\mathbf{B} \\ \mathbf{A}(i_1,:) = \mathbf{B}(i_2,:)}}{\sub{\mathbf{A}^{-1}(i_1,:)}{\emptyset}{p_{i_1,i_2}}} \right]
\end{equation*}
and 
\begin{equation*}
p_{i_1,i_2} = |\mathbf{A}^{-1}(i_1,:)| - \max\bigl(0,|\mathbf{A}^{-1}(i_1,:)|-|\mathbf{B}^{-1}(i_2,:)|\bigr)
\end{equation*}
This removes from $\mathbf{A}$ as many copies of a row as are present in $\mathbf{B}$ (up to all of the copies of that row in $\mathbf{A}$).
In SQL syntax, it is
\begin{equation*}
\text{\sf SELECT $\ast$ FROM $\mathbf{A}$ EXCEPT SELECT $\ast$ FROM $\mathbf{B}$}
\end{equation*}
\end{definition}

\noindent The use of $\pi_1$, projection onto the first coordinate, in the above expression of $S$, is intended to correct for the fact that the set
\[
  \sub{\mathbf{A}^{-1}(i_1,:)}{\emptyset}{p_{i_1,i_2}}
\]
is technically a subset of $I_\mathbf{A} \times \{1\}$. Taking $\pi_1$ makes $S$ a subset of $I_\mathbf{A}$ instead.

The notion of $\Sub{A}{B}{n}$ allows for the notion of a set difference of $\mathbf{A}$ and $\mathbf{B}$ to be defined as well, being defined as in Definition~\ref{set difference definition} except with $p_{i_1,i_2}=0$.

The notion of $\Sub{A}{B}{n}$ also allows for all duplicate elements of an associative array to be removed, effectively allowing for set semantics, by taking
\begin{equation*}
\operatorname{Set}(\mathbf{A}) = \mathbb{I}_S \mathbf{A}
\end{equation*}
where
\begin{equation*}
S= \pi_1 \left[ \bigcup_{i \in I_\mathbf{A}}{\Sub{\mathbf{A}^{-1}(i,:)}{\emptyset}{1}}\right]
\end{equation*}

\subsection{Operations Involving Functions of Certain Entries in a Row}

\begin{definition}
Suppose $J$ is a set of column keys. If $\mathbf{A}$ is an array, then the \emph{$J$-column indexed entries of a row} $\mathbf{A}(i,:)$ are the entries $\mathbf{A}(i,j)$ where $j\in J$.
\end{definition}

\begin{definition}[Select] \label{select definition}
Suppose $\varphi$ is a boolean-valued function (so taking values in $\{0,1\}\subset \mathbb{V}$) of the $J$-column indexed entries of a row, whose values we denote as $\varphi(\mathbf{A}(k_1,J))$. Then we define the \emph{select operation} (determined by $\varphi$) by
\begin{equation*}
\sigma_{\varphi(J)}(\mathbf{A}) = \bigl[ [ \varphi(\mathbf{A}(:,J)) ~
\varphi(\mathbf{A}(:,J))^\intercal] \otimes \mathbb{I}_{I_\mathbf{A}}\bigr] ~
\mathbf{A}
\end{equation*}
where $\varphi(\mathbf{A}(:,J))$ is the column vector
\begin{equation*}
\varphi(\mathbf{A}(:,J)) = \kbordermatrix{ & 1 \\ i_1 & \varphi(\mathbf{A}(i_1,J)) \\ \vdots & \vdots \\ i_n & \varphi(\mathbf{A}(i_n,J)) }
\end{equation*}
This operation selects those rows of $\mathbf{A}$ which evaluate to true under $\varphi$.
In SQL syntax, it is
\begin{equation*}
\text{\sf SELECT $\ast$ FROM $\mathbf{A}$ where $\varphi(\mathbf{A}_{\cdot J(1),\ldots, J(n)})$}
\end{equation*}
\end{definition}

\begin{definition}[Theta Join] \label{theta join definition}
Suppose $\theta$ is a boolean-valued function on the $J_1$-column-indexed entries of a first row and the $J_2$-column-indexed entries of a second row.  Then define the \emph{theta join operation} (determined by $\theta$) by
\begin{align*}
\mathbf{A} \bowtie_{\theta(J_1,J_2)} \mathbf{B} & = \sigma_{\theta(J_1,J_2)}\Bigl( [\mathbf{A}
\mathbin{\text{\textcircled{$\otimes$}}}
\mathbbm{1}_{K_3,\{1\}}] \\
& \quad \quad \oplus \rho_{\{2\}\times K_4, K_4\times\{2\},f} [\mathbbm{1}_{K_1,\{2\}}
\mathbin{\text{\textcircled{$\otimes$}}}
\mathbf{B}]\Bigr)
\end{align*}
where $f: (2,k) \mapsto (k,2)$. 

\noindent This operation selects pairs of rows from $\mathbf{A}$ and $\mathbf{B}$ which evaluate to true under $\theta$.
In SQL syntax, it is
\begin{align*}
\text{\sf SELECT $\ast$ FROM $\mathbf{A},\mathbf{B}$ WHERE} \\
\theta(\mathbf{A}_{\cdot J_1(1),\ldots, J_1(n)}, \mathbf{B}_{\cdot J_2(1),\ldots, J_2(n)})
\end{align*}
\end{definition}

\noindent The theta join operation creates new column indices for the resulting rows by ``tagging'' them with $1$ and $2$; this ensures that there is no conflict between them when performing the array addition.  

If needed, it can be assumed that whenever a theta join operation is performed, the non-zero column indices of the first and second array are distinct, and the column indices $K_2 \times \{1\}$ and $K_4 \times \{2\}$ can be identified with the corresponding column indices of $K_2$ and $K_4$, respectively.

Dealing with the case where the non-zero column indices of the two arrays are not necessarily distinct, it can be required that $\theta$ evaluates to true exactly when the values at those indices agree and are defined. To achieve this, the array addition $\oplus$ can be replaced with a new operation $\oplus_=$ for which
\begin{equation*}
v \oplus_= w = \begin{cases} v & \text{if $w=0$} \\ w & \text{if $v=0$} \\ v & \text{if $v=w$} \\ \mathrm{undefined} & \text{otherwise} \end{cases}
\end{equation*}
(This ``undefined'' value only shows up to be removed upon use of the selection $\sigma_{\theta(J_1,J_2)}$, so there is no need not worry about the effect of it on the algebra.)

\begin{definition}[Extended Projection] \label{extended projection definition}
Suppose $\varphi$ is a function of the $J$-column indexed entries of a row and $j'$ is a column key.  Define the \emph{extended projection} (determined by $\varphi$ and $j'$) by
\begin{align*}
{}_{j'}\Pi_{\varphi(J)}(\mathbf{A}) & = \rho_{\{1\},\{j'\}}\bigl( \varphi(\Pi_J(\mathbf{A})(:,J)) \bigr) \\
& = \kbordermatrix{ & j' \\ i_1 & \varphi(\mathbf{A}(i_1,J)) \\ \vdots & \vdots \\ i_n & \varphi(\mathbf{A}(i_n,J)) }
\end{align*}
and $I_\mathbf{A} = \{i_1,\ldots, i_n\}$.  This replaces the $J$-indexed columns with a single column $j'$ whose entries are computed by $\varphi$.
In SQL syntax, it is
\begin{equation*}
\text{\sf SELECT $\varphi(\mathbf{A}_{\cdot J(1),\ldots, J(n)})$ AS $j'$ FROM $\mathbf{A}$}
\end{equation*}
\end{definition}

\noindent Taking liberties with the notation, it is typical to write
\[
  \varphi(\Pi_J(\mathbf{A})(:,J))
\]
as 
\begin{equation*}
\varphi(\Pi_J(\mathbf{A})(:,J)) = \Pi_J(\mathbf{A}) \arrayprod{\varphi}{\otimes} \mathbbm{1}_{J,\{j'\}}
\end{equation*}
since the entries are computed as 
\begin{equation*}
\varphi(\mathbf{A}(i,j_1),\ldots, \mathbf{A}(i,j_m)) = \varphi\bigl( \mathbf{A}(i,j_1)\otimes 1,\ldots, \mathbf{A}(i,j_m)\otimes 1\bigr)
\end{equation*}
which is remarkably close to
\[
  \bigoplus_{j\in J}{(\mathbf{A}(i,j)\otimes 1)}
\]  
In fact, if $\varphi$ is an iterated (commutative, associative) binary operation $\ast$, then this is the same as
\[
  \Pi_J(\mathbf{A}) \arrayprod{\ast}{\otimes} \mathbbm{1}_{J,\{j'\}}
\]

\begin{definition}[Aggregation] \label{aggregation definition}
Suppose $j$ and $j'$ are column keys and $f$ is a function of finitely-supported tuples of elements in $\mathbb{V}$ (i.e. all but finitely-many elements are $0$) taking values in $\mathbb{V}$.  Define the \emph{aggregation} (determined by $j,j'$ and $f$) by
\begin{equation*}
{}_{j}\mathcal{G}_{f(j')}(\mathbf{A}) = \mathbf{P} \arrayprod{f}{\otimes} \mathbf{A}
\end{equation*}
\begin{equation*}
= \kbordermatrix{ & 1 \\ i_1 & f\left( \mathbf{P}(i_1,i_1)\otimes \mathbf{A}(i_1,j'), \ldots, \mathbf{P}(i_1,i_n) \otimes \mathbf{A}(i_n,j')\right) \\ \vdots & \vdots \\ i_n & f\left( \mathbf{P}(i_n,i_1) \otimes \mathbf{A}(i_1,j'), \ldots, \mathbf{P}(i_n,i_n) \otimes \mathbf{A}(i_n,j') \right) }
\end{equation*}
where 
\begin{equation*}
\mathbf{P} =  [\mathbb{I}_{I_\mathbf{A}} \arrayprod{\oplus}{\otimes} \mathbf{A}(:,j)] \arrayprod{\oplus}{\delta} [\mathbb{I}_{I_\mathbf{A}} \arrayprod{\oplus}{\otimes} \mathbf{A}(:,j)]^\intercal
\end{equation*}
This operation applies the function $f$ (the \emph{aggregate function}) on all the values of column $j'$ in $\mathbf{A}$ that share a common value in column $j$.
In SQL syntax, it is
\begin{equation*}
\text{\sf SELECT $f_{j'}$ FROM $\mathbf{A}$ GROUP BY $j$}
\end{equation*}
\end{definition}

\noindent Unlike in the cases of selection, theta join, and extended projection, where the domains of the relevant functions ($\varphi$ in the case of selection and extended projection, $\theta$ in the case of theta join) are explicitly given, the domain of $f$ is not explicitly given. (We've said it is a function of finitely-supported tuples of elements in $\mathbb{V}$, but without restricting the possible indices, there are too many of these to even form a set.)

In all practical considerations, the set of possible column keys can be assumed to be finite; in this case, consider all finitely-supported tuples of elements in $\mathbb{V}$ indexed by those possible column keys.

Another practical consideration is that $f$ is \emph{symmetric}, in that permuting those indexing column keys does not affect the result. In this case, take $f$ to simply be a function of any finite multiset of non-zero values in $\mathbb{V}$, passing on the set-theoretic difficulties to that of multisets.

If the values $\mathbf{A}(i,j)$ and $\mathbf{A}(i',j)$ are equal and non-zero, then the aggregate ${}_j\mathcal{G}_{f(j')}(\mathbf{A})$ will also have its $i$-th and $i'$-th entries equal. Moreover, the row keys of the aggregate are the same as those of $\mathbf{A}$ (at least, the non-zero rows).

By additionally (array) multiplying $\mathbb{I}_{I_\mathbf{A},\mathbb{V},f}^\intercal$ on the left, where $f(i) = \mathbf{A}(i,j)$, every row of the aggregate represents unique information with row keys equal to the value $\mathbf{A}(i,j)$ that was used to select the values being aggregated.

Finally, to use a column key $j''$ in place of the default $1$, (array) multiply $\mathbb{I}_{\{1\},\{j''\}}$ on the right.

\section{Properties of Relational Algebra Operations}

Since relations are defined by associative arrays with either strong or weak equivalence, to ensure that these operations are defined on relations, they must be invariant under strong and weak equivalence.

\begin{prop}
Each operation is invariant under strong equivalence.
Each operation (except for multiset difference) is invariant under weak equivalence.
\end{prop}

\noindent The fact that multiset difference (Definition~\ref{set difference definition}) is not invariant under weak equivalence is not a random occurrence -- this is due to the fact that the definition of multiset difference seeks to remove only a certain number of instances of a row. If, instead, every instance of a row was removed, then this new operation would be invariant under both strong and weak equivalence.

Many of the desirable properties of each of the relational algebra operations can be proven using array algebra with the above definitions of those operations.

\begin{prop}
\mbox{}
\begin{enumerate}
\item If there is a fixed set $K$ of column keys, then $\Pi_{K}$ acts as the identity map. 
\item $\Pi_\emptyset$ sends every array to the zero array $\mathbs{0}$.
\item $\rho_{J/J,{\mathrm{id}}_J}$ acts as the identity map.  
\item $\mathbs{0}$ is an identity under $\cup$.
\item $\mathbs{0}$ is an annihilator under $\cap$.
\item $\mathbs{0}$ is a right identity and left annihilator under $\setminus$.
\item If $\varphi \equiv 1$, then $\sigma_\varphi$ acts as the identity map.
\item If $\varphi \equiv 0$, then $\sigma_\varphi$ sends every array to $\mathbs{0}$.
\item If $\theta \equiv 0$, then $\mathbf{A} \bowtie_{\theta(J,J')} \mathbf{B} = \mathbs{0}$.
\item If $J= \{j'\}$ and $\varphi(v)=v$, then ${}_{j'}\Pi_{\varphi(J)}$ acts as the identity map.  If $\varphi \equiv 0$, then ${}_{j'}\Pi_{\varphi(J)}$ sends every array to $\mathbs{0}$.
\item If $J_1,J_2$ are sets of column indices, then
\[
  \Pi_{J_1} \circ \Pi_{J_2} = \Pi_{J_1 \cap J_2}
\] 
where $\circ$ is function composition.

\item $\Pi_J$ preserves $\cup,\cap,\setminus$.

\item If $J_1,J_2,J_3,J_4$ are sets of column indices and
\[
  f: J_1 \to J_2
\]
and
\[
  g: J_3 \to J_4
\]
are bijections, then it need not be the case that
\[
  \rho_{J_1/J_2,f} \circ \rho_{J_3/J_4,g} = \rho_{J_3/J_4,g} \circ \rho_{J_1/J_2,f}
\]
even up to strong or weak equivalence, where $\circ$ is function composition.

\item $\rho_{J_1/J_2,f}$ preserves $\cup,\cap,\setminus$.

\item Both $\cup$ and $\cap$ are commutative and associative (up to both weak and strong equivalence, and up to a canonical renaming of column keys).

\item Both $\cup$ and $\cap$ distribute over one-another (up to both weak and strong equivalence, and up to a canonical renaming of column keys).

\item $\setminus$ is neither commutative nor associative (even up to strong or weak equivalence).

\end{enumerate}

\end{prop}

The proofs of all of the above proposition is beyond the space limitations of this work.  However, they are straightforward given the definitions.  As an example proof using array algebra, consider the proof that
\[
  \Pi_{J_1} \circ \Pi_{J_2} = \Pi_{J_1 \cap J_2}
\]
and that
\[
  \Pi_J(\mathbf{A} \cup \mathbf{B}) = \Pi_J(\mathbf{A}) \cup \Pi_J(\mathbf{B})
\]

\begin{proof}
\item By Definition~\ref{project definition},
\begin{align*}
\Pi_{J_1}(\Pi_{J_2}(\mathbf{A}))
& = (\mathbf{A} ~ \mathbb{I}_{J_2}) ~ \mathbb{I}_{J_1} \\
& = (\mathbf{A} \arrayprod{\oplus}{\otimes} \mathbb{I}_{J_2}) \arrayprod{\oplus}{\otimes} \mathbb{I}_{J_1} \\
& = \mathbf{A} \arrayprod{\oplus}{\otimes} (\mathbb{I}_{J_2} \arrayprod{\oplus}{\otimes} \mathbb{I}_{J_1})
\end{align*}
Thus, it suffices to show that
\[
  \mathbb{I}_{J_2} \arrayprod{\oplus}{\otimes} \mathbb{I}_{J_1} = \mathbb{I}_{J_1 \cap J_2}
\]
By Definition~\ref{array multiply definition},
\begin{equation*}
\left( \mathbb{I}_{J_2} \arrayprod{\oplus}{\otimes} \mathbb{I}_{J_1} \right)(i,j) = \bigoplus_{k \in J_1 \cup J_2}{\mathbb{I}_{J_2}(i,k) \otimes \mathbb{I}_{J_1}(k,j)}
\end{equation*}
The term
\[
  \mathbb{I}_{J_2}(i,k) \otimes \mathbb{I}_{J_1}(k,j)
\]
is $1$ if and only if $i=k \in J_2$ and $k=j \in J_1$, and $0$ otherwise.  This only occurs when $i=k=j \in J_1 \cap J_2$, and this contributes the only possible non-zero term of the sum.  This shows that
\begin{align*}
(\mathbb{I}_{J_2} \arrayprod{\oplus}{\otimes} \mathbb{I}_{J_1})(i,j) & = \begin{cases} 1 & \text{if $i=j \in J_1 \cap J_2$} \\ 0 & \text{otherwise} \end{cases} \\
& = \mathbb{I}_{J_1\cap J_2}(i,j)
\end{align*}

For preservation of union, Definition~\ref{union definition} gives
\begin{align*}
\Pi_J(\mathbf{A} \cup \mathbf{B})
& = \Pi_J\bigl((\mathbb{I}_{I_\mathbf{A}\times \{1\},I_\mathbf{A}} ~
\mathbf{A})
& \oplus & ~~ (\mathbb{I}_{I_\mathbf{B}\times\{2\},I_\mathbf{B}} ~
\mathbf{B})\bigr) \\
& = \bigl( (\mathbb{I}_{I_\mathbf{A}\times \{1\},I_\mathbf{A}} ~
\mathbf{A})
& \oplus & ~~ (\mathbb{I}_{I_\mathbf{B}\times\{2\},I_\mathbf{B}}
\mathbf{B})\bigr) ~ 
\mathbb{I}_J \\
& = ((\mathbb{I}_{I_\mathbf{A}\times \{1\},I_\mathbf{A}} ~
\mathbf{A}) ~
\mathbb{I}_J)
& \oplus & ~~ ((\mathbb{I}_{I_\mathbf{B}\times\{2\},I_\mathbf{B}} ~
\mathbf{B}) ~
\mathbb{I}_J ) \\
& = (\mathbb{I}_{I_\mathbf{A}\times \{1\},I_\mathbf{A}} ~
(\mathbf{A} ~
\mathbb{I}_J) ) 
& \oplus & ~~(\mathbb{I}_{I_\mathbf{B}\times \{2\},I_\mathbf{B}} ~
(\mathbf{B} ~
\mathbb{I}_J) ) \\
& = \bigl(\mathbb{I}_{I_\mathbf{A}\times\{1\},I_\mathbf{A}} 
\Pi_J(\mathbf{A})\bigr) 
& \oplus & ~~ \bigl(\mathbb{I}_{I_\mathbf{B}\times\{2\},I_\mathbf{B}} ~
\Pi_J(\mathbf{B})\bigr)
\end{align*}
Now, this is nearly in the form $\Pi_J(\mathbf{A}) \cup \Pi_J(\mathbf{B})$, with the only issue being that wherever a $I_\mathbf{A}$ or $I_\mathbf{B}$ show up, there should instead be $I_{\Pi_J(\mathbf{A})}$ or $I_{\Pi_J(\mathbf{B})}$, respectively.  However, this replacement can be made due to the fact that taking a projection can only make $I_\mathbf{A}$ (resp. $I_\mathbf{B}$) smaller in size. Indeed, if $f: J_1 \pfun J_2$ is a partial injection, $J_3 \subset J_1$, and $I_\mathbf{C} \subset f[J_3] = \{f(j) \mid j\in J_3\}$, then 
\begin{equation*}
\mathbb{I}_{J_1,J_2,f} ~
\mathbf{C} = \mathbb{I}_{J_3,J_2,f|_{J_3}} ~
\mathbf{C}
\end{equation*}

\end{proof}

One benefit of proving these properties of the relational algebra operations as defined via the array algebra operations is that it gives a better understanding of how these operations work without resorting to equality up to strong or weak equivalence; in many cases, the relation is outright equality, or at least equality up to strong or weak equivalence where the relabeling is in some sense ``canonical''.

Another benefit is in performance \cite{kepner2016}; thanks to the fact that associativity, commutativity, and distributivity of $\oplus,\otimes,\arrayprod{\oplus}{\otimes}$  lead to performance increases since the relational algebra operations can be built up by the associative algebra operations.

\section{Conclusion}

SQL, NoSQL, and NewSQL databases are specialized to deal with certain domains, and all three can be useful in a single context. For this reason, polystore databases have been developed to bridge these three concepts. 

Associative arrays provide a mathematical framework through which the mathematical cores of SQL, NoSQL, and NewSQL can be reduced, allowing for polystore databases like BigDAWG to translate between the three data types inherent to these databases -- sets (SQL), graphs (NoSQL), and matrices (NewSQL). 

Future work will focus on exploring additional properties that the associative array perspective provides with regards to relational algebra, providing analysis of optimizations, and the potential application of quantifying uncertainly in database queries.


\section*{Acknowledgment}
%
%

The authors wish to acknowledge the following individuals for their contributions: Michael Stonebraker, Sam Madden, Bill Howe, David Maier, Alan Edelman, Dave Martinez, Sterling Foster, Paul Burkhardt, Victor Roytburd, Bill Arcand, Bill Bergeron, David Bestor, Chansup Byun, Mike Houle, Matt Hubbell, Mike Jones, Anna Klein, Pete Michaleas, Lauren Milechin, Julie Mullen, Andy Prout, Tony Rosa, Sid Samsi, and Chuck Yee.



\bibliographystyle{IEEEtran}
\bibliography{aarabib}
%

\end{document}